\documentclass[12pt]{article}
\usepackage{epsf, cite, amsmath, amssymb}
\usepackage{epsfig}

\makeatletter
\renewcommand\section{\@startsection {section}{1}{\z@}%
                                   {-3.5ex \@plus -1ex \@minus -.2ex}
                                   {2.3ex \@plus.2ex}%
                                   {\normalfont\large\bfseries}}
\renewcommand\subsection{\@startsection{subsection}{2}{\z@}%
                                     {-3.25ex\@plus -1ex \@minus -.2ex}%
                                     {1.5ex \@plus .2ex}%
                                     {\normalfont\bfseries}}
\makeatother

\let\non\nonumber

\newcommand{\bea}{\begin{eqnarray}}
\newcommand{\eea}{\end{eqnarray}}
\newcommand{\be}{\begin{equation}}
\newcommand{\ee}{\end{equation}}

\newcommand{\hlf}{\frac{1}{2}}

\newcommand{\T}{\theta}
\newcommand{\G}{\Gamma}

\newcommand{\e}{\epsilon}

\newcommand{\LL}{\Lambda}
\newcommand{\dd}{\delta}

\newcommand{\rr}{\rightarrow}
\newcommand{\m}{\mu}
\newcommand{\n}{\nu}
\newcommand{\s}{\sigma}
\newcommand{\p}{\partial}

\newcommand{\al}{\alpha}

\newcommand{\rt}{\tilde R}
 
\newcommand{\C}[1]{$(\ref{#1})$}

\begin{document}
\begin{titlepage}

\begin{center}

June 19, 2003
\hfill                  hep-th/0306193

\hfill EFI-03-23

\vskip 2 cm
{\Large \bf The UV/IR Interplay in Theories with   
\vskip 0.1in
Space-time Varying Non-commutativity}\\
\vskip 1.25 cm {Daniel Robbins\footnote{email: robbins@theory.uchicago.edu} and
Savdeep Sethi\footnote{email: sethi@theory.uchicago.edu} 
}\\
{\vskip 0.75cm
Enrico Fermi Institute, University of Chicago,
Chicago, IL
60637, USA\\}

\end{center}

\vskip 2 cm

\begin{abstract}
\baselineskip=18pt

We consider scalar field theory with space and space-time-dependent 
non-commuta-tivity.
In perturbation theory, we find that the structure
of the UV/IR mixing is quite different from cases with constant
non-commutativity. In particular, UV/IR mixing becomes
intertwined in an interesting way with violations of momentum
conservation. 


\end{abstract}

\end{titlepage}

\pagestyle{plain}
\baselineskip=19pt
\section{Introduction}

Recently there have been several examples, coming from string theory, of
non-commutative structure with a space-time-dependent noncommutativity
parameter~\cite{Hashimoto:2002nr,Dolan:2002px,Cai:2002sv,Dasgupta:2003us,
Lowe:2003qy,
Cerchiai:2003yu}.  In many of these examples, the non-commuting
directions in space-time form a nilpotent
Lie algebra:
\be \label{lie}
\left[\hat x^\m,\hat x^\n\right] = i\T_\rho^{\m\n}\hat x^\rho,
\ee
where the structure constants $\T_\rho^{\m\n}$ really are constant.
We will use hats to denote non-commuting coordinates, and unhatted variables
for their commutative counterparts.
This algebra is very special because the non-commutativity depends
only linearly on
the space-time coordinates.

This kind of  non-commutative
space-time has been considered before; see, for
example,~\cite{Majid:1994cy, Madore:2000en, Imai:2000kq,
  Agostini:2003vg}.
 One can use a Weyl
quantization procedure to write down a Groenewald-Moyal star product 
\cite{Madore:2000en, Kathotia}.  To write the explicit form of the star-product
requires the Baker-Campbell-Hausdorff formula.  The examples that interest 
us will have either all double commutators or all triple commutators vanishing,
and we can write explicitly
\be \label{double}
f\ast g(x) = e^{\frac{i}{2}\T_\rho^{\m\n}x^\rho\p_\m^{(1)}\p_\n^{(2)}}f(x_1)
g(x_2)|_{x_1=x_2=x},
\ee
when all double commutators vanish, or
\be \label{triple}
f\ast g(x) = e^{\frac{i}{2}\T_\rho^{\m\n}x^\rho\p_\m^{(1)}\p_\n^{(2)}+
\frac{1}{12}\T_\lambda^{\m\n}\T_\rho^{\lambda\s}x^\rho\p_\m^{(1)}\p_\n^{(2)}
(\p_\s^{(1)}-\p_\s^{(2)})}f(x_1)g(x_2)|_{x_1=x_2=x},
\ee
when all triple commutators vanish. 

This product is automatically associative because the associativity
condition on the product (for general $\T^{\m\n}$ parameters)
\be
\T^{\m\n}\p_\n\T^{\rho\s}+\T^{\rho\n}\p_\n\T^{\s\m}+\T^{\s\n}\p_\n\T^{\m\rho}
=0
\ee
becomes simply the Jacobi identity on the structure coefficients
\be
\T_\lambda^{\m\n}\T_\n^{\rho\s}+\T_\lambda^{\rho\n}\T_\n^{\s\m}+
\T_\lambda^{\s\n}\T_\n^{\m\rho}=0,
\ee
which is guaranteed  by the Lie algebra structure~\C{lie}. 

Quantum field
theories on these non-commutative spaces are quite fascinating, and
largely unexplored (though see~\cite{Imai:2000kq}\ for a similar study on another Lie 
algebra example). Our aim in this
work is to study the perturbative structure of these theories along the
lines of~\cite{Filk:1996dm, Minwalla:1999px}. As we shall see, the
physics is quite different from theories with constant
non-commutativity. We will consider scalar field theory. We will show
that space-time-dependent
non-commutativity explicitly breaks momentum conservation in a way
that becomes entwined with UV/IR mixing. We would like to interpret
our results
using a `dipole-like' explanation which proved useful in the case of
constant non-commutativity~\cite{Sheikh-Jabbari:1999vm,
  Bigatti:1999iz}. However, it seems likely that such an explanation
will involve strings rather than particles since (in a stringy
context) $H=dB$ is non-zero. This, in turn, may require an analysis along
the lines described in~\cite{Cornalba:2001sm}.

There are many directions to explore: a basic issue is the perturbative consistency of 
these theories. It should be possible to address this issue using the techniques 
of~\cite{Gomis:2000zz}. When the non-commutativity parameter depends on time, the resulting 
non-local theory is quite unusual (even if the parameter is purely spatial). Understanding the 
conditions under which such theories make sense is important.  
The case of Yang-Mills
theory built from a space-time varying product is also interesting, but
more subtle  even in its classical
definition. Yang-Mills theories with space-time varying
non-commutativity appear in holographic descriptions of
cosmological space-times~\cite{Hashimoto:2002nr}, and have also been
considered recently for phenomenology~\cite{Calmet:2003jv}.

\vskip 0.25 in

{\bf Note added:} After we had completed this project, we received an interesting paper~\cite{BG}\ 
which contains related observations.

\section{Two Particular Examples}
We will focus on two examples that are both realized in string
theory. 

\subsection{A space-dependent case}

The first example is from \cite{Lowe:2003qy}, and is realized in
massive type IIA. The structure constants are,
\be\label{mIIA}
\T_3^{12} = -\T_3^{21} = \al, 
\ee
corresponding to a non-commutative space with relations
\be
\left[\hat x^1,\hat x^2 \right] = i \al \hat x^3,\qquad \left[\hat
  x^1,\hat x^3 \right] = \left[\hat x^2,\hat x^3 \right] = 0. 
\ee
The star product of several functions is given by
\be \label{mIIAp}
f_1\ast f_2\ast\cdots\ast f_n = e^{\frac{i}{2}\al x^3\sum_{a<b}(\p_1^{(a)}
\p_2^{(b)}-\p_2^{(a)}\p_1^{(b)})}f_1f_2\cdots f_n.
\ee
An important property satisfied by this product is the relation
\be
\label{totaldiv}
f\ast g = fg + \mathrm{total\ divergence}.
\ee
In this case, this is true for the same reason as in the case of the
standard star product. Namely, consider a term in the integrated product
\be
\int{ {1\over n!} \left\{ \frac{i}{2}\T_\rho^{\m\n}x^\rho
  \p_\m^{(1)}\p_\n^{(2)} \right\}^n f(x_1) g(x_2)|_{x_1=x_2=x}.}
\ee
With the choice \C{mIIA}, we can freely integrate by parts to move one
$\p_\m$ from $f$ to $g$. This gives zero up to a total
divergence. This relation means that the tree-level propagator for an
action involving this star product will agree with the usual commutative
case. Note that because of the explicit $x^3$-dependence in the
product, the momentum charge $P^3$ is not conserved.

\subsection{A space-time-dependent case}
\label{NBalg}

In the case of the algebra coming from study of the null-brane
quotient~\cite{Figueroa-O'Farrill:2001nx}, 
originally described in \cite{Hashimoto:2002nr}, the structure constants are
\be
\T_+^{xz} = -\T_+^{zx} = \T_x^{-z} = -\T_x^{z-} = \tilde R,
\ee
which corresponds to a non-commutative space-time with non-vanishing relations
\be
\left[\hat x,\hat z \right] = i \tilde R \, \hat x^+,\qquad \left[\hat
  x^-,\hat z \right] = i \tilde R \, \hat x. 
\ee

As derived in \cite{Cerchiai:2003yu}, this algebra leads to a closed
form Groenewald-Moyal
star product between functions of the commutative variables 
\bea\label{longtriple}
f_1\ast f_2 &=& \exp\left[\frac{i}{2}\T_\rho^{\m\n}x^\rho\p_\m^{(1)}\p_\n^{(2)}
-\frac{1}{12}\T_\rho^{\m\n}\T_\n^{\s\lambda}x^\rho\p_\s^{(1)}\p_\lambda^{(2)}
(\p_\m^{(1)}-\p_\m^{(2)})\right]f_1f_2 \non\\
&=& \exp\left[\frac{i}{2}\tilde Rx^+\left(\p_x^{(1)}\p_z^{(2)}-\p_z^{(1)}
\p_x^{(2)}\right)+\frac{i}{2}\tilde Rx\left(\p_-^{(1)}\p_z^{(2)}-\p_z^{(1)}
\p_-^{(2)}\right)\right.\\
&& \left.+\frac{1}{12}\tilde R^2x^+\left(\p_{-z}^{(1)}\p_z^{(2)}-
\p_{zz}^{(1)}\p_-^{(2)}-\p_-^{(1)}\p_{zz}^{(2)}+\p_z^{(1)}\p_{-z}^{(2)}\right)
\right]f_1f_2.\non
\eea
Here $\p^{(i)}$ is understood to act only on $f_i$ (so in particular no
derivatives act on the coordinates in the exponent).  It is less
trivial to check that
\be
f\ast g = fg + \mathrm{total\ derivative}
\ee
in this case. To see this, we need to examine terms in the expansion
of \C{longtriple}\ 
\bea\label{tripleterms} && \int {1\over n!} \left[\frac{i}{2}\tilde Rx^+\left(\p_x^{(1)}\p_z^{(2)}-\p_z^{(1)}
\p_x^{(2)}\right)+\frac{i}{2}\tilde Rx\left(\p_-^{(1)}\p_z^{(2)}-\p_z^{(1)}
\p_-^{(2)}\right)\right.\\
&& \left.+\frac{1}{12}\tilde R^2x^+\left(\p_{-z}^{(1)}\p_z^{(2)}-
\p_{zz}^{(1)}\p_-^{(2)}-\p_-^{(1)}\p_{zz}^{(2)}+\p_z^{(1)}\p_{-z}^{(2)}\right)
\right]^n f_1f_2.\non 
\eea
There are no explicit factors of $z, x^-$ in \C{tripleterms}\ so we can
integrate by parts to make the triple product terms, and the double
product terms proportional to $x$ vanish (up to total
divergences). That leaves the first term in \C{tripleterms}\ which can
now safely be integrated by parts to give zero. As in the previous
example, quadratic terms in an action built from this star product
reduce to those of a commutative theory.

At first glance, one might worry that the term in \C{longtriple}\ that is quadratic in
$\tilde R$ is not a phase, since there is no $i$ in front of it.  However, if
write the interaction in terms of $x^3$ and the $z, x^-$ momenta
then the partial derivatives are replaced by factors of $ik_\m$, and all the
terms become phases.

It is not very difficult to generalize the result \C{longtriple}\ to a finite product
$f_1\ast f_2\ast\cdots\ast f_n$.  The result takes the deceptively simple form
\be
\hat A\hat B\hat C f_1\cdots f_n,
\ee
where $\hat A$, $\hat B$, $\hat C$ are mutually commuting differential 
operators taking the explicit forms
\bea
\label{hatops}
\hat A &=& \prod_{a<b}\exp\left[\frac{i}{2}\tilde Rx^+\left(\p_x^{(a)}
\p_z^{(b)}-\p_z^{(a)}\p_x^{(b)}\right)+\frac{i}{2}\tilde Rx\left(\p_-^{(a)}
\p_z^{(b)}-\p_z^{(a)}\p_-^{(b)}\right)\right]\non\\
\hat B &=& \prod_{a<b}\exp\left[\frac{1}{12}\tilde R^2x^+\left(\p_{-z}^{(a)}
\p_z^{(b)}-\p_{zz}^{(a)}\p_-^{(b)}-\p_-^{(a)}\p_{zz}^{(b)}+\p_z^{(a)}
\p_{-z}^{(b)}\right)\right]\\
\hat C &=& \prod_{a<b<c}\exp\left[\frac{1}{6}\tilde R^2x^+\left(2\p_z^{(a)}
\p_-^{(b)}\p_z^{(c)}-\p_-^{(a)}\p_z^{(b)}\p_z^{(c)}-
\p_z^{(a)}\p_z^{(b)}\p_-^{(c)}\right)\right]. \non
\eea
This gives us the modification to the vertex factor in $\phi^n$ scalar field
theory in position space.  Compare this with the usual constant $\T$
case, where there is only an operator corresponding to $\hat A$, and it has no 
explicit position dependence.  So in momentum space, it can simply be rewritten
as a a phase factor,
\be
V(k_1,\ldots,k_n) = \prod_{a<b}e^{-\frac{i}{2}\T^{\m\n}k_\m^{(a)}k_\n^{(b)}},
\ee
and the only modification to the momentum space Feynman rules is the inclusion
of this factor at each vertex.  Our situation is a little more complicated
because of the appearance of both positions and derivatives in our phases.
In particular, because of the $x$ and $x^+$-dependence, interactions will
not conserve the $P^x, P^-$ momenta.

\section{Perturbation Theory for the Space-Dependent Case}
\subsection{Feynman rules}

We would now like to consider a scalar field theory Lagrangian constructed from
the space-dependent product \C{mIIAp}, 
\be
\mathcal{L} = \hlf\left(\p_\m\phi\right)^2 + \hlf m^2\phi^2 + \frac{g^2}{n!}
\underbrace{\phi\ast\phi\ast\cdots\ast\phi}_{\mathrm{n\ times}}.
\ee
This theory is a little simpler than the time-dependent case, and so
provides a good set-up for us to first study the effects of space-time
varying non-commutativity. Since this theory is static, we can
analytically continue to the Euclidean 
theory.  Because of \C{totaldiv}, the quadratic terms in the action
involve only ordinary 
products.  The Feynman rules can then be written down in the following mixed 
picture (i.e. with integrations both over the positions of the vertices and 
over the momenta of the internal lines). We will simplify these rules
in a moment. The rules are:
\begin{enumerate}
\item For each external line, a factor $e^{-ip\cdot x}$.
\item For each internal line, a factor 
$\frac{1}{p^2+m^2+i\e}e^{-ip\cdot (x-y)}$.
\item For each vertex, a factor $g^2e^{-\frac{i}{2}\al x^3\sum_{a<b}(p_1^{(a)}
p_2^{(b)}-p_2^{(a)}p_1^{(b)})}$.
\item An integration $\int d^4x$ over the position of each vertex.
\item An integration $\int\frac{d^4p}{(2\pi)^4}$ over each internal momentum.
\end{enumerate}

We can perform the integrations over $x^0$, $x^1$, and $x^2$ to produce
delta functions which enforce conservation of $p_0$, $p_1$, and $p_2$ at each
vertex.  We could also perform the integration over $x^3$ to produce a delta
function of the form
\be \label{momnon}
\dd\left(\sum_a p_3^{(a)}+\frac{\al}{2}\sum_{a<b}(p_1^{(a)}p_2^{(b)}-
p_2^{(a)}p_1^{(b)})\right).
\ee
This already makes clear the difference between this case and the case
of constant
non-commutativity. With constant non-commutativity, the
analogue of \C{momnon}\ is a phase of the form
\be \label{vertex} \dd\left(\sum_a p_3^{(a)} \right)
\exp\left\{i\frac{\al}{2}\sum_{a<b}(
p_1^{(a)}p_2^{(b)}-p_2^{(a)}p_1^{(b)})\right\}, \ee
while for us, the phase is replaced by momentum non-conservation. 
However, for some purposes, we will find it more helpful to leave the
$x^3$ integrations
undone for now. Our simplified Feynman rules become:
\begin{enumerate}
\item For each internal line, a factor $\frac{1}{p^2+m^2+i\e}$.
\item For each vertex, a factor 
\bea 
&& g^2\exp\left\{-ix^3\left[\sum_a p_3^{(a)}+\hlf\al\sum_{a<b}(p_1^{(a)}p_2^{(b)}-
p_2^{(a)}p_1^{(b)})\right]\right\} \times \cr && (2\pi)^3\dd(\sum_a p_0^{(a)})
\dd(\sum_a p_1^{(a)})\dd(\sum_a p_2^{(a)}).
\eea
\item An integration $\int dx^3$ over the $x^3$ position of each vertex.
\item An integration $\int\frac{d^4p}{(2\pi)^4}$ over each internal momentum.
\end{enumerate}

\subsection{$\phi^4$ in four dimensions}

Let us now specialize to the non-commutative $\phi^4$ theory in four dimensions.
We will compute the one-loop corrections to the propagator of this
theory. In conventional non-commutative field theory, the vertex
modification \C{vertex}\ is invariant under cyclic permutation of the
momenta~\cite{Filk:1996dm}. This result relies
only on
momentum conservation in the non-commutative directions,  and so it
continues to hold for us. 

\begin{figure}
\begin{center}
\resizebox{0.8\textwidth}{!}{\includegraphics{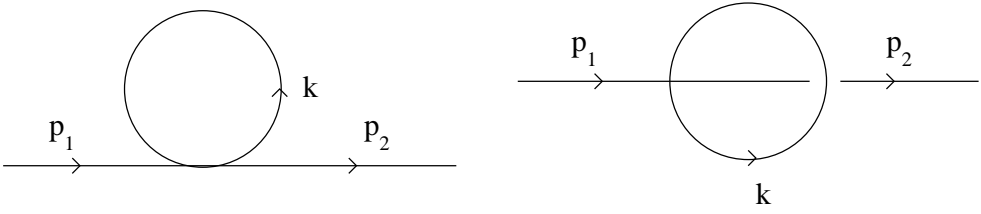}}
\end{center}
\caption{The planar and non-planar one-loop contributions to the $\phi^4$
propagator.}
\label{oneloop}
\end{figure}

The same comment applies to the
decomposition into planar and non-planar diagrams, again because
momenta in the non-commuting directions are conserved at each vertex. 
As in conventional non-commutative $\phi^4$ theory, there are then two
diagrams that contribute:
one planar diagram and one non-planar diagram, both of which are
depicted in Figure \ref{oneloop}.

In the planar diagram, the extra phase factors at the vertex cancel and we have
a contribution
\bea
\G_{1\ \mathrm{planar}}^{(2)} &=& \frac{g^2}{3(2\pi)^5}\dd^3\left(p_\perp^{(1)}
-p_\perp^{(2)}\right)\int dx^3\int\frac{d^4k}{k^2+m^2}e^{-ix^3(p_3^{(1)}-
p_3^{(2)})} \non\\
&=& \frac{g^2}{3(2\pi)^4}\dd^4\left(p^{(1)}-p^{(2)}\right)\int
\frac{d^4k}{k^2+m^2},
\eea
where $p_\perp$ represents the three conserved components of momentum, i.e. 
$p_0$, $p_1$, and $p_2$.  This is exactly the same as in commutative 
$\phi^4$ theory up to a symmetry factor. This diagram is quadratically
divergent. 

The non-planar diagram is more interesting.  Here the extra phase factors do
not cancel and we get a contribution
\be
\G_{1\ \mathrm{np}}^{(2)} = \frac{g^2}{6(2\pi)^5}\dd^3\left(
p_\perp^{(1)}-p_\perp^{(2)}\right)\int dx^3\int\frac{d^4k}{k^2+m^2}
e^{-ix^3 \{ p_3^{(1)}-p_3^{(2)}-\al(p_1^{(1)}k_2-p_2^{(1)}k_1)\} }.
\ee
As usual for a massive theory, the only divergences come from the
region of large $k$.  IR divergences are supressed.

To examine this result more closely, we will rewrite the propagator using
an integral over a Schwinger parameter 
\be
\frac{1}{k^2+m^2} = \int_0^\infty ds e^{-s(k^2+m^2)}.
\ee
The $k$ integrations then become Gaussians, and on performing those integrals, we
obtain
\bea
\G_{1\ \mathrm{planar}}^{(2)} &=& \frac{g^2}{48\pi^2}\dd^4\left(p^{(1)}-
p^{(2)}\right)\int_0^\infty\frac{ds}{s^2}e^{-s m^2} \\
\G_{1\ \mathrm{np}}^{(2)} &=& \frac{g^2}{192\pi^3}\dd^3\left(
p_\perp^{(1)}-p_\perp^{(2)}\right)\int_0^\infty\frac{ds}{s^2}\int dx\,
e^{-s m^2-\frac{\al^2}{4s}p_{nc}^2 (x^3)^2-ix^3\Delta p_3}.  \label{regulated}
\eea
We have used the abbreviations $p_{nc}^2 = (p_1^{(1)})^2 + (p_2^{(1)})^2$ and
$\Delta p_3 = p_3^{(1)} - p_3^{(2)}$ in the second integral.  The large $k$
divergences have now become divergences at small values of $s$, so to
regulate these integrals we include a multiplicative factor of 
$\exp[-1/(\LL^2 s)]$.  This gives us
\bea
\G_{1\ \mathrm{planar}}^{(2)} &=& \frac{g^2}{48\pi^2}\dd^4\left(p^{(1)}-
p^{(2)}\right)\int_0^\infty\frac{ds}{s^2}e^{-s m^2-
\frac{1}{\LL^2 s}} \\
\G_{1\ \mathrm{np}}^{(2)} &=& \frac{g^2}{192\pi^3}\dd^3\left(
p_\perp^{(1)}-p_\perp^{(2)}\right)\int_0^\infty\frac{ds}{s^2}\int dx^3\,
e^{-s m^2-\frac{\al^2}{4s}p_{nc}^2 (x^3)^2-ix^3\Delta
  p_3-\frac{1}{\LL^2 s}}.
\eea

Since the $x^3$ integral in the non-planar amplitude above is now Gaussian, we
will certainly be able to that integral, but first let us note a couple
of features.  For fixed values of $x^3$, the integral above consists of a phase
factor times a standard Schwinger integral with an effective
space-dependent cutoff
\be
\LL_{eff}^2 = \frac{1}{1/\LL^2+\al^2 (x^3)^2 p_{nc}^2/4}.
\ee
This is the same result that one would obtain from the noncommutative theory 
with constant $\T^{12} = \al x^3$.  Also, the fact that the $x^3$ integral is 
Gaussian implies the interaction is localized in the $x^3$ 
direction for $p_{nc}^2 \neq 0$.  The non-planar interaction is heavily
supressed for large $|x^3|$.

We can see UV/IR mixing quite nicely from \C{regulated}. If we
first take the non-commutative momenta $p_{nc}^2 \rightarrow 0$ then we can integrate
over $x^3$ to get
$$ \delta (\Delta p_3) $$
and a quadratically divergent contribution. In other words, we recover
the commutative result. However, if $p_{nc}^2 \neq 0$ then (setting $\Delta
p_3=0$ for simplicity), we can integrate out $x^3$ to get a
$s\rightarrow 0$  divergence
$$ \sim \int { ds \over s^{3/2}}$$ 
which is softer than the planar contribution. The order in which we take 
$p_{nc}^2\rightarrow 0$ and $ \LL \rightarrow \infty$ matters just
as in~\cite{Minwalla:1999px}. 

Let us be more precise. Performing the $x^3$ integration with $p_{nc}^2\neq
0$, we are left with
\be
\G_{1\ \mathrm{np}}^{(2)} = \frac{g^2}{192\pi^3}\dd^3\left(
p_\perp^{(1)}-p_\perp^{(2)}\right)\sqrt\frac{4\pi}{\al^2 p_{nc}^2}\int_0^\infty
\frac{ds}{s^{3/2}}e^{-s m_{eff}^2-\frac{1}{\LL^2 s}},
\ee
where we have now defined an effective mass
\be
m_{eff}^2 = m^2+\frac{(\Delta p_3)^2}{\al^2 p_{nc}^2}.
\ee
Apart from the momentum-dependent factor out front, this is the type of
Schwinger integral that we would evaluate for a three-dimensional theory.  In
fact, the integral can be evaluated exactly using the results listed in
Appendix \ref{integrals}.  The result, along with the
conventional planar contribution, is 
\bea
\label{phi4oneloop}
\G_{1\ \mathrm{planar}}^{(2)} &=& \frac{g^2}{48\pi^2}\dd^4\left(p^{(1)}-
p^{(2)}\right)\left(\LL^2-m^2\ln(\frac{\LL^2}{m^2})+O(1)\right) \non\\
\G_{1\ \mathrm{np}}^{(2)} &=& \frac{g^2}{96\pi^2}\dd^3\left(
p_\perp^{(1)}-p_\perp^{(2)}\right)\frac{\LL}{\al\sqrt{p_{nc}^2}}\,e^{-2
\sqrt{m_{eff}^2/\LL^2}}.
\eea

We can now compare this exact result for the non-planar amplitude
\C{phi4oneloop}\ with our qualitative
expectations. First, only the combination
$\al^2p_{nc}^2$ appears.  As $\al^2 p_{nc}^2\rr\infty$, i.e., either for infinite 
non-commutativity and generic momenta, or for finite non-commutativity and very
large external momenta in the noncommuting directions, the non-planar amplitude
is heavily suppressed relative to the planar amplitude.  This is analogous to
the large non-commutativity limit in theories with constant $\T$.  For finite
non-zero $\al^2 p_{nc}^2$, the UV divergence of the amplitude is softened from
quadratic to linear in $\LL$ -- the divergence that one would expect
for $\phi^4$ theory in three dimensions. As we will see in the space-time non-commutative 
case, as $\theta$ depends on more coordinates, the UV divergence for the non-planar diagram 
becomes softer.    

Finally, consider the limit $\al^2p_{nc}^2\rr 0$.  For $\Delta p_3$ nonzero,
$m_{eff}^2\rr 0$ and the exponential causes the amplitude to vanish.  For
$\Delta p_3=0$, the amplitude diverges.  Indeed, since
\be
\int d(\Delta p_3)\frac{1}{\al\sqrt{p_{nc}^2}}\,e^{-2\sqrt{m_{eff}^2/\LL^2}}
= \LL\int du\,e^{-2\sqrt{u^2+m^2/\LL^2}} < \infty,
\ee
it follows that in this limit the non-planar amplitude is simply proportional to
$\dd(\Delta p_3)$ as we expect:
\bea
\lim_{\al^2p_{nc}^2\rr 0}\G_{1\ \mathrm{np}}^{(2)} &=& \frac{g^2}{96\pi^2}
\dd^4\left(p^{(1)}-p^{(2)}\right)\LL^2\int du\,e^{-2\sqrt{u^2+m^2/\LL^2}} 
\non\\
&\simeq& \frac{g^2}{96\pi^2}\dd^4\left(p^{(1)}-p^{(2)}\right)\left(\LL^2-
m^2\ln(\frac{\LL^2}{m^2})+O(1)\right).
\eea
So in this limit the correct commutative behavior is recovered.  Note also that
for finite $\al$ there are always external momenta ($p_1, p_2\approx 1/\al$)
for which the non-planar contribution is non-neglible when compared to
the planar diagram. For completeness, we include in Appendix \ref{sixdim}\ the analogous 
1-loop computation for $\phi^3$ theory in six dimensions.

\section{The Space-time-dependent Case}

We now turn to the algebra of section \C{NBalg}\ derived from the
null-brane quotient.  We can again write
down Feynman rules for $\phi^4$ theory in four dimensions defined using the
star-product \C{longtriple}.  The Minkowski space rules are
\begin{enumerate}
\item For each internal line, a factor $\frac{-i}{p^2+m^2+i\e}$.
\item For each vertex, a factor 
\begin{multline}
g^2\exp\left\{-ix\left[\sum_a p_x^{(a)}+\hlf\rt\sum_{a<b}(p_-^{(a)}p_z^{(b)}-
p_z^{(a)}p_-^{(b)})\right]-ix^+\left[\sum_a p_+^{(a)}\right.\right.\\
\left.\left.+\hlf\rt\sum_{a<b}
(p_x^{(a)}p_z^{(b)}-p_z^{(a)}p_x^{(b)})-\frac{\rt^2}{6}\sum_{a<b<c}\left(
2p_z^{(a)}p_-^{(b)}p_z^{(c)}-p_-^{(a)}p_z^{(b)}p_z^{(c)}-p_z^{(a)}p_z^{(b)}
p_-^{(c)}\right)\right]\right\}\\
\times(2\pi)^2\dd(\sum_a p_-^{(a)})\dd(\sum_a p_z^{(a)}).
\end{multline}
\item An integration $\int dx^+dx$ over the position of each vertex.
\item An integration $\int\frac{d^4p}{(2\pi)^4}$ over each internal momentum.
\end{enumerate}
Notice that the vertex factor contribution that would have come from $\hat B$
in \C{hatops} is absent.  One can check that it in fact vanishes by
conservation of $p_-$ and $p_z$.

Let us again consider the one-loop correction to the propagator.  As before we
have contributions from the same two diagrams displayed in figure \ref{oneloop}.  
{}For the planar 
diagram, the extra vertex factors cancel, and we are again left with the same 
contribution as in the commutative case.

The non-planar contribution is now given by
\bea
& & \G_{1\ \mathrm{np}}^{(2)} = \frac{g^2}{6(2\pi)^6}\dd\left(
p_-^{(1)}-p_-^{(2)}\right)\dd\left(p_z^{(1)}-p_z^{(2)}\right)\int dx^+dx
\int d^4k\frac{-i}{k^2+m^2+i\e}\\
&&\times\exp\left[-ix\left(\Delta p_x+\rt(p_zk_--p_-k_z)\right)-ix^+\left(
\Delta p_++\rt(p_zk_x-p_x^{(2)}k_z-\hlf p_z\Delta p_x)\right.\right.\non\\
&&\left.\left.-\hlf\rt^2(p_zp_zk_--p_zp_-k_z-p_zk_zk_-+p_-k_zk_z)\right)
\right].\non
\eea
Because the (bare) propagator is unmodified, we see that the
pole structure in the above integrals is the same as the usual case.  This makes it plausible 
that we can rotate the $k_0$ contour, replacing
$k_0$ with $i\tilde k_0$ (here we are using $k_\pm = (k_0\pm k_1)/\sqrt 2$).
If we go ahead and make this change, and then replace the propagator with a
Schwinger integral, we obtain (for simplicity, we drop the tildes on $k_0$)
\bea
\G_{1\ \mathrm{np}}^{(2)} &=& \frac{g^2}{6(2\pi)^6}\dd\left(
p_-^{(1)}-p_-^{(2)}\right)\dd\left(p_z^{(1)}-p_z^{(2)}\right)\int_0^\infty 
ds\int dx^+dxd^4k\exp\left[\vphantom{\frac{1}{\sqrt 2}}-s\left(k^2+
m^2\right)\right.\non\\
&&-ix\left(\Delta p_x-\rt p_-k_z-\frac{1}{\sqrt 2}\,\rt p_zk_1\right)
-ix^+\left(\Delta p_++\rt(p_zk_x-p_x^{(2)}k_z-\hlf p_z\Delta p_x)\right.\non\\
&&\left.+\hlf\rt^2(\frac{1}{\sqrt 2}\,p_zp_zk_1+p_zp_-k_z-
\frac{1}{\sqrt 2}\,p_zk_zk_1-p_-k_zk_z)\right)\non\\
&&\left.+\frac{1}{\sqrt 2}\,x\rt p_zk_0-\frac{1}{2\sqrt 2}\,x^+
\rt^2p_z(p_z-k_z)k_0\right].
\eea

The momentum integrals are now Gaussian.  Performing them, we obtain
\bea
\G_{1\ \mathrm{np}}^{(2)} &=& \frac{g^2}{3\cdot 2^7\pi^4}\dd\left(
p_-^{(1)}-p_-^{(2)}\right)\dd\left(p_z^{(1)}-p_z^{(2)}\right)\int_0^\infty 
ds\int dx^+dx\frac{1}{s^{3/2}B^{1/2}}\\
&&\times\exp\left[-s m^2-ix\Delta p_x-ix^+(\Delta p_+-
\hlf\rt p_z\Delta p_x)-\frac{\rt^2p_z^2}{4s}(x^+)^2\right.\non\\
&&\left.-\frac{\rt^2p_-^2}{4B}(x+\frac{p_x^{(2)}}{p_-}x^+-\hlf\rt p_zx^+)^2
\right],\non
\eea
where we have defined $B=s-\frac{i}{2}\rt^2p_-x^+$.

If we now perform the integral over $x$, followed by the integral over $x^+$,
we obtain
\bea
\G_{1\ \mathrm{np}}^{(2)} &=& \frac{g^2}{96\pi^3}\dd\left(
p_-^{(1)}-p_-^{(2)}\right)\dd\left(p_z^{(1)}-p_z^{(2)}\right)
\frac{1}{\rt^2|p_-p_z|}\int_0^\infty \frac{ds}{s}e^{-s m_{eff}^2
-\frac{1}{s\LL^2}}\non\\
&=& \frac{g^2}{48\pi^3}\dd(\Delta p_-)\dd(\Delta p_z)\frac{1}{\rt^2|p_-p_z|}
K_0\left(2\frac{m_{eff}}{\LL}\right),
\eea
where now
\be
m_{eff}^2 = m^2+\frac{(\Delta p_x)^2}{\rt^2p_-^2}+\frac{1}{\rt^2p_z^2}
\left(\Delta p_+-\frac{\bar p_x\Delta p_x}{p_-}\right)^2,
\ee
and $\bar p_x=\hlf(p_x^{(1)}+p_x^{(2)})$.  $K_0$ is a modified Bessel function
of the second kind.

For nonzero $\rt$, $p_-$, and $p_z$, the large $\LL$ behaviour is
\be
\G_{1\ \mathrm{np}}^{(2)} = \frac{g^2}{96\pi^3}\dd\left(
p_-^{(1)}-p_-^{(2)}\right)\dd\left(p_z^{(1)}-p_z^{(2)}\right)
\frac{1}{\rt^2|p_-p_z|}\ln\left(\frac{\LL}{m_{eff}}\right).
\ee
So for nonzero $\rt$, $p_-$, and $p_z$, the expected UV divergence is softened
from quadratic to logarithmic.  We should also note that in the limit
$\rt\rr\infty$, the non-planar amplitude is again suppressed (for generic 
momenta) relative to the planar amplitude.

Finally, let us consider the behaviour for IR values of the external momenta.
It turns out that as we take either $p_z\rr 0$ or $p_-\rr 0$, we restore one
delta function, while if we take $\rt\rr 0$ (or equivalently both $p_-\rr 0$
and $p_z\rr 0$ at once) we restore the commutative limit and full momentum
conservation.  Explicitly,
\bea
\lim_{p_z\rr 0}\G_{1\ \mathrm{np}}^{(2)} &=& \frac{g^2}{96\pi^2}\dd(\Delta p_-)
\dd(\Delta p_z)\dd(\Delta p_+ - \frac{\bar p_x\Delta p_x}{p_-})
\frac{\LL}{\rt|p_-|}e^{-\frac{2}{\LL}\sqrt{m^2+
\frac{(\Delta p_x)^2}{\rt^2p_-^2}}}, \non\\
\lim_{p_-\rr 0}\G_{1\ \mathrm{np}}^{(2)} &=& \frac{g^2}{96\pi^2}\dd(\Delta p_-)
\dd(\Delta p_z)\dd(\Delta p_x)\frac{\LL}{\rt\sqrt{p_z^2+\bar p_x^2}}
e^{-\frac{2}{\LL}\sqrt{m^2+\frac{(\Delta p_+)^2}{\rt^2(p_z^2+\bar p_x^2)}}},
\non\\
\lim_{\rt\rr 0}\G_{1\ \mathrm{np}}^{(2)} &=& \frac{g^2}{48\pi^2}\dd^4(\Delta p)
m\LL K_1\left(2\frac{m}{\LL}\right).
\eea
We should stress, in closing, that this theory is quite unusual. Energy and momentum are not conserved, 
and as a consequence, we expect to observe strange decay phenomena. However, the existence of strange 
theories of this kind is suggested by the behavior of strings in cosmological backgrounds.

\section*{Acknowledgements}
We would like to thank Aki Hashimoto and Vikram Divvuri for 
helpful discussions. S.~S. would like to thank the organizers of the 2003 Amsterdam Workshop on 
String Theory and Quantum Gravity for their hospitality.  
The work of D.~R. is supported
in part by a Julie Payette--NSERC PGS B Research Scholarship, and by  
NSF CAREER Grant No. PHY-0094328.
The work of S.~S. is supported in part by NSF CAREER Grant No.
PHY-0094328, and by the Alfred P. Sloan Foundation.

\appendix
\section{Some Useful Integrals}
\label{integrals}
We note that most of the integrals over Schwinger parameters that we need
to perform give modified Bessel functions of the second kind, since
\be
I_\m\left(m,\LL\right)\equiv\int_0^\infty ds\,s^{-\m}
\exp\left[-s m^2-\frac{1}{s\LL^2}\right] = 
2\left(m\LL\right)^{\m-1}K_{\m-1}\left(2\frac{m}{\LL}\right).
\ee
Specifically, we need
\bea
I_2\left(m,\LL\right) &=& 2m\LL K_1\left(2\frac{m}{\LL}\right) \simeq \LL^2
- m^2\ln\left(\frac{\LL^2}{m^2}\right),\non\\
I_{3/2}\left(m,\LL\right) &=& 2\sqrt{m\LL}K_{1/2}\left(2\frac{m}{\LL}\right) =
\sqrt\pi\LL e^{-2\frac{m}{\LL}}, \\
I_1\left(m,\LL\right) &=& 2K_0\left(2\frac{m}{\LL}\right) \simeq 
\hlf\ln\left(\frac{\LL}{m}\right).\non
\eea
We have included the exact closed form for $K_{1/2}$, and the large $\LL$ 
expansions for $K_0$ and $K_1$.

\section{The One-loop Propagator in Six Dimensions}
\label{sixdim}

We can repeat the one-loop analysis for $\phi^4$ theory in the case of
 $\phi^3$
theory in six dimensions.  The Feynman rules are basically the same, and we
again have one planar and one non-planar diagram to compute. 

The loop diagrams involve two interaction vertices so we need to introduce two Schwinger parameters. 
Since the computations are straightforward, we will only display the results in terms of 
these parameters.
\bea
\G_{1\ \mathrm{planar}}^{(2)} &=& \frac{g^4}{256\pi^3}\dd^6\left(p^{(1)}-
p^{(2)}\right)\int ds_1ds_2\frac{1}{\left(s_1+s_2\right)^2}
\frac{1}{\sqrt{(s_1+s_2)^2+\frac{1}{4}s_1s_2\al^2p_{nc}^2}} \non\\
&& \times\exp\left[-\left(s_1+s_2\right)^2m^2-
\frac{s_1s_2}{s_1+s_2}p_\perp^2-
\frac{s_1s_2}{s_1+s_2+
\frac{s_1s_2\al^2p_{nc}^2}{4(s_1+s_2)}}p_3^2\right] \non\\
\G_{1\ \mathrm{np}}^{(2)} &=& \frac{g^4}{256\pi^{7/2}}\dd^5\left(
p_\perp^{(1)}-p_\perp^{(2)}\right)\frac{1}{\al\sqrt{p_{nc}^2}}\int ds_1ds_2
\frac{1}{\left(s_1+s_2\right)^{5/2}} \\
&& \times\exp\left[-\left(s_1+s_2\right)^2m_{eff}^2-
\frac{s_1s_2}{s_1+s_2}\bar p^2\right] \non
\eea
Here we have defined the quantities, 
\bea
p_{nc}^2 &=& (p_1^{(1)})^2 + (p_2^{(1)})^2, \non\\
m_{eff}^2 &=& m^2 + \frac{(\Delta p_3)^2}{\al^2p_{nc}^2}, \non\\
\Delta p_3 &=& p_3^{(1)}-p_3^{(2)}, \non \\
\bar p^2 &=& (p_\perp^{(1)})^2+\frac{1}{4}\left(p_3^{(1)}+p_3^{(2)}\right)^2.
\non
\eea


\begin{thebibliography}{10}

\bibitem{Hashimoto:2002nr}
A.~Hashimoto and S.~Sethi, ``Holography and string dynamics in time-dependent
  backgrounds,'' {\em Phys. Rev. Lett.} {\bf 89} (2002) 261601,
\href{http://www.arXiv.org/abs/hep-th/0208126}{{\tt hep-th/0208126}}.

\bibitem{Dolan:2002px}
L.~Dolan and C.~R. Nappi, ``Noncommutativity in a time-dependent background,''
  {\em Phys. Lett.} {\bf B551} (2003) 369--377,
\href{http://www.arXiv.org/abs/hep-th/0210030}{{\tt hep-th/0210030}}.

\bibitem{Cai:2002sv}
R.~G.~Cai, J.~X.~Lu and N.~Ohta, ``NCOS and D-branes in time-dependent 
backgrounds,'' {\em Phys. Lett.} {\bf B551} (2003) 178--186,
\href{http://www.arXiv.org/abs/hep-th/0210206}{{\tt hep-th/0210206}}.

\bibitem{Dasgupta:2003us}
K.~Dasgupta, G.~Rajesh, D.~Robbins, and S.~Sethi, ``Time-dependent warping,
  fluxes, and NCYM,'' {\em JHEP} {\bf 03} (2003) 041,
\href{http://www.arXiv.org/abs/hep-th/0302049}{{\tt hep-th/0302049}}.

\bibitem{Lowe:2003qy}
D.~A. Lowe, H.~Nastase, and S.~Ramgoolam, ``Massive IIA string theory and
  Matrix theory compactification,''
\href{http://www.arXiv.org/abs/hep-th/0303173}{{\tt hep-th/0303173}}.

\bibitem{Cerchiai:2003yu}
B.~L. Cerchiai, ``The Seiberg-Witten map for a time-dependent background,''
\href{http://www.arXiv.org/abs/hep-th/0304030}{{\tt hep-th/0304030}}.

\bibitem{Majid:1994cy}
S.~Majid and H.~Ruegg, ``Bicrossproduct structure of kappa Poincare group 
and noncommutative geometry,'' {\em Phys. Lett.} {\bf B334} (1994) 348--354,
\href{http://www.arXiv.org/abs/hep-th/9405107}{{\tt hep-th/9405107}}.

\bibitem{Madore:2000en}
J.~Madore, S.~Schraml, P.~Schupp, and J.~Wess, ``Gauge theory on noncommutative
  spaces,'' {\em Eur. Phys. J.} {\bf C16} (2000) 161--167,
\href{http://www.arXiv.org/abs/hep-th/0001203}{{\tt hep-th/0001203}}.

\bibitem{Imai:2000kq}
S.~Imai and N.~Sasakura, ``Scalar field theories in a Lorentz-invariant
three-dimensional  noncommutative space-time,'' {\em JHEP} {\bf 09} (2000)
032, \href{http://www.arXiv.org/abs/hep-th/0005178}{{\tt hep-th/0005178}}.

\bibitem{Agostini:2003vg}
A.~Agostini, G.~Amelino-Camelia and F.~D'Andrea, ``Hopf-algebra 
description of noncommutative-spacetime symmetries,''
\href{http://www.arXiv.org/abs/hep-th/0306013}{{\tt hep-th/0306013}}.

\bibitem{Kathotia}
V.~Kathotia, ``Kontsevich's Universal Formula for Deformation Quantization and
  the Campbell-Baker-Hausdorff Formula, I,''
  \href{http://www.arXiv.org/abs/math.QA/9811174}{{\tt math.QA/9811174}}.

\bibitem{Filk:1996dm}
T.~Filk, ``Divergencies in a field theory on quantum space,'' {\em Phys. Lett.}
  {\bf B376} (1996)
53--58.

\bibitem{Minwalla:1999px}
S.~Minwalla, M.~Van~Raamsdonk, and N.~Seiberg, ``Noncommutative perturbative
  dynamics,'' {\em JHEP} {\bf 02} (2000) 020,
\href{http://www.arXiv.org/abs/hep-th/9912072}{{\tt hep-th/9912072}}.

\bibitem{Sheikh-Jabbari:1999vm}
M.~M. Sheikh-Jabbari, ``Open strings in a B-field background as electric
  dipoles,'' {\em Phys. Lett.} {\bf B455} (1999) 129--134,
\href{http://www.arXiv.org/abs/hep-th/9901080}{{\tt hep-th/9901080}}.

\bibitem{Bigatti:1999iz}
D.~Bigatti and L.~Susskind, ``Magnetic fields, branes and noncommutative
  geometry,'' {\em Phys. Rev.} {\bf D62} (2000) 066004,
\href{http://www.arXiv.org/abs/hep-th/9908056}{{\tt hep-th/9908056}}.

\bibitem{Cornalba:2001sm}
L.~Cornalba and R.~Schiappa, ``Nonassociative star product deformations 
for D-brane worldvolumes in  curved backgrounds,'' {\em Commun. Math. Phys.}
{\bf 225} (2002) 33--66,
\href{http://www.arXiv.org/abs/hep-th/0101219}{{\tt hep-th/0101219}}.

\bibitem{Gomis:2000zz}
J.~Gomis and T.~Mehen, ``Space-time noncommutative field theories and
  unitarity,'' {\em Nucl. Phys.} {\bf B591} (2000) 265--276,
\href{http://www.arXiv.org/abs/hep-th/0005129}{{\tt hep-th/0005129}}.

\bibitem{Calmet:2003jv}
X.~Calmet and M.~Wohlgenannt, ``Effective field theories on non-commutative
  space-time,''
\href{http://www.arXiv.org/abs/hep-ph/0305027}{{\tt hep-ph/0305027}}.

\bibitem{BG}
O.~Bertolami and L.~Guisado, ``Noncommutative field theory and violation of
  translation invariance,'' \href{http://www.arXiv.org/abs/hep-th/0306176}{{\tt
  hep-th/0306176}}.

\bibitem{Figueroa-O'Farrill:2001nx}
J.~Figueroa-O'Farrill and J.~Simon, ``Generalized supersymmetric fluxbranes,''
  {\em JHEP} {\bf 12} (2001) 011,
\href{http://www.arXiv.org/abs/hep-th/0110170}{{\tt hep-th/0110170}}.

\end{thebibliography}


\providecommand{\href}[2]{#2}\begingroup\raggedright\endgroup


\end{document}